\documentclass[prl,aps,twocolumn,superscriptaddress,showpacs]{revtex4-1}

\usepackage{graphicx}% Include figure files
\usepackage{color}%
\usepackage{epstopdf}

\begin{document}

\title{Structural and magnetic characterisation of iron oxyselenides Ce$_{2}$O$_{2}$Fe$_{2}$OSe$_{2}$ and Nd$_{2}$O$_{2}$Fe$_{2}$OSe$_{2}$ }

\author{E. E. McCabe}
\affiliation{School of Physical Sciences, University of Kent, Canterbury, CT2 7NH, UK}
\affiliation{Department of Chemistry, Durham University, Durham DH1 3LE, UK}
\author{A. S. Wills}
\affiliation{Department of Chemistry, University College London, 20 Gordon Street, London, WC1H 0AJ, UK}
\author{L. Chapon}
\affiliation{Diffraction Group, Institute Laue Langevin, 71 avenue des Martyrs, 38000 Grenoble, France}
\affiliation{ISIS Facility, Rutherford Appleton Labs, Chilton Didcot, OX11 0QX, UK}
\author{P. Manuel}
\affiliation{ISIS Facility, Rutherford Appleton Labs, Chilton Didcot, OX11 0QX, UK}
\author{J. S. O. Evans}
\affiliation{Department of Chemistry, Durham University, Durham DH1 3LE, UK}
\date{\today}

\begin{abstract}

We present here an investigation of the magnetic ordering in the Mott insulating oxyselenide materials $Ln_{2}$O$_{2}$Fe$_{2}$OSe$_{2}$ ($Ln$ = Ce, Nd). Neutron powder diffraction data are consistent with a non-collinear multi-$k$ ordering on the iron sublattice structure and analysis indicates a reduced magnetic correlation length perpendicular to the [Fe$_{2}$O]$^{2+}$ layers. The magnetic role of the $Ln^{3+}$ cations is investigated and Ce$^{3+}$ moments are found to order for T $\leq$ 16 K.

\end{abstract}

\maketitle

\section{1. Introduction}

The discovery of iron-based superconductivity \cite{Kamihara-2008} in several mixed anion systems has driven research to understand the magnetism of these materials which is related to their superconducting behaviour \cite{Wysocki-2011}. The 1111 family of iron-based superconductors derive from $Ln$FeAsO ($Ln$ = trivalent lanthanide) and adopt the ZrCuSiAs-structure \cite{Johnson-1974}, composed of alternating layers of anti-fluorite-like edge-linked FeAs$_{4}$ tetrahedra alternating with fluorite-like edge-linked O$Ln_{4}$ tetrahedra. Although the superconductivity arises in the iron arsenide layers, the $Ln^{3+}$ cation has a significant role in both tuning the superconducting $T_{\mathrm{c}}$ in the doped materials \cite{Kamihara-2008, Ren-2008} and in defining the magnetism of the parent non-superconducting materials (such as the spin-reorientations in NdFeAsO \cite{Tian-2010} and PrFeAsO \cite{Bhoi-2011}). This has led to much research into the interplay between transition metal and lanthanide magnetic sublattices in these and related mixed-anion systems \cite{Chi-2008, Gornostaeva-2013, Jesche-2009, Kimber-2010, Lee-2012, Maeter-2009, Marcinkova-2012, Marcinkova-2010, Ni-2011, Tsukamoto-2011, Zhang-2013, McCabe-2014-Ce2O2FeSe2, McGuire-2009}.

\begin{figure*}[t] 
\includegraphics[width=0.8\linewidth,angle=0.0]{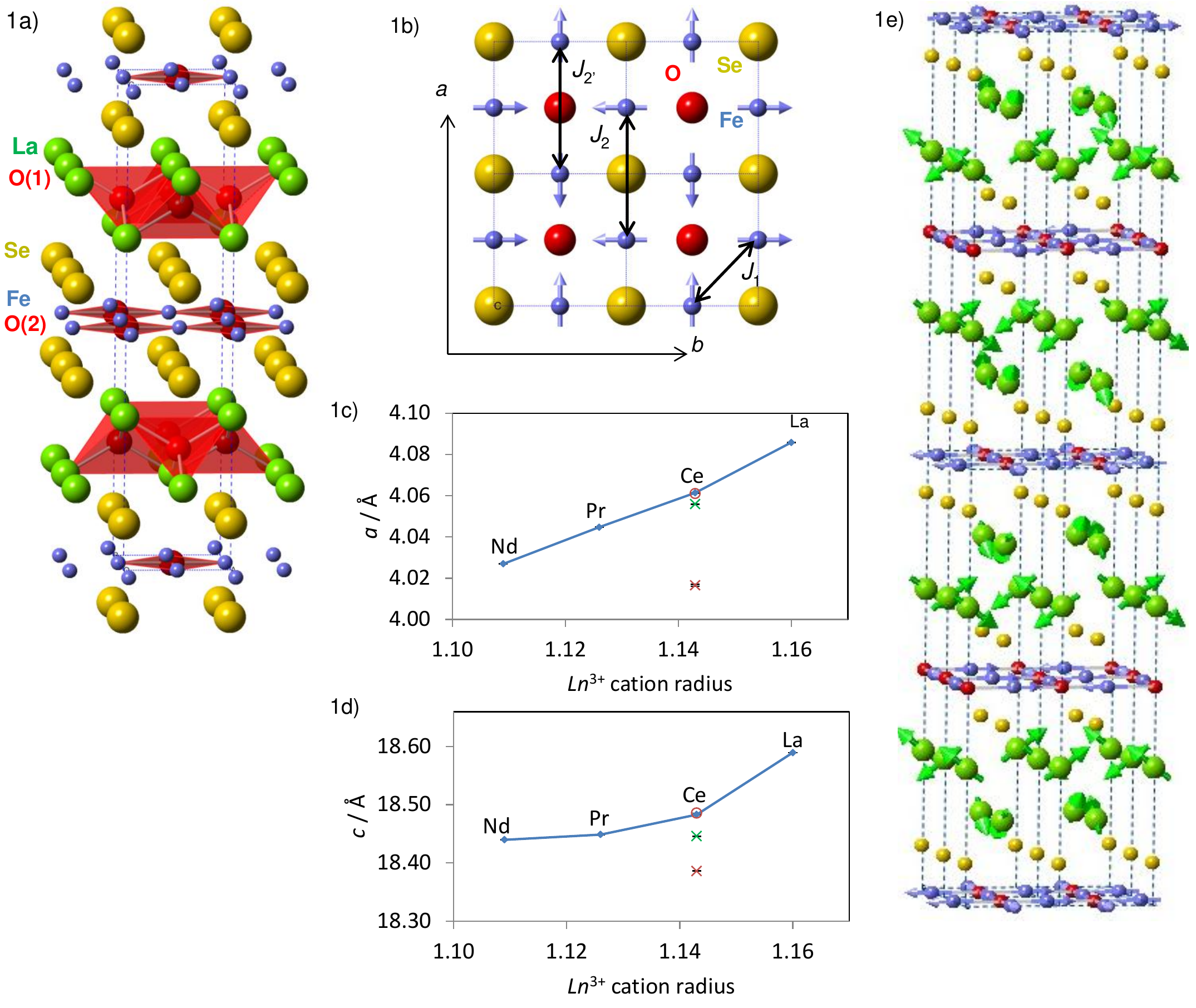}
\caption{[color online]  a) nuclear cell of $Ln_{2}$O$_{2}$Fe$_{2}$OSe$_{2}$ with O$^{2-}$, Se$^{2-}$, Fe$^{2+}$ and $Ln^{3+}$ ions in red, yellow, blue and green, respectively, showing [$Ln_{2}$O$_{2}$]$^{2+}$ fluorite-like layers built from edge-shared O$Ln_{4}$ tetrahedra in red; b) shows the in-plane arrangement of magnetic moments for the 2-$k$ model, with O$^{2-}$ (red) Se$^{2-}$ (yellow) and Fe$^{2+}$ (blue) ions shown, and Fe$^{2+}$ moments shown by blue arrows; the three intraplanar exchange interactions $J_{1}$, $J_{2}$ and $J_{2'}$ are also shown; c) and d) show plots of unit cell parameters $a$ and $c$ for $Ln_{2}$O$_{2}$Fe$_{2}$OSe$_{2}$ determined from Rietveld analysis using room temperature XRPD data (data in blue are described in this work including quench-cooled Ce$_{2}$O$_{2}$Fe$_{2}$OSe$_{2}$); data in red are determined for the main ($\bigcirc$) and secondary ($\times$) phases in two phase samples of ``Ce$_{2}$O$_{2}$Fe$_{2}$OSe$_{2}$"; green points are those reported for Ce$_{2}$O$_{2}$Fe$_{2}$OSe$_{2}$ by Ni et al \cite{Ni-2011}); e) illustrates the magnetic ordering of Ce (green) and Fe (blue) moments (O(1) sites are omitted for clarity).}
\label{structures}
\end{figure*}

La$_{2}$O$_{2}$Fe$_{2}$OSe$_{2}$ was the first reported member of the ``$M_{2}$O" ($M$ = transition metal) family of oxyselenide materials and adopts a layered structure with fluorite-like [La$_{2}$O$_{2}$]$^{2+}$ slabs (analogous to those in 1111 iron oxyarsenides) separated by Se$^{2-}$ anions from [Fe$_{2}$O]$^{2+}$ sheets (Figure \ref{structures}) \cite{Mayer-1992}. The Fe$^{2+}$ cations are coordinated by two O(2) ions within these anti-CuO$_{2}$ sheets, and by four Se$^{2-}$ ions above and below the sheets, forming a network of face-shared FeO$_{2}$Se$_{4}$ trans octahedra. Much research has been carried out to understand the exchange interactions and magnetic ordering present in these [Fe$_{2}$O]$^{2+}$ sheets \cite{Kabbour-2008}. La$_{2}$O$_{2}$Fe$_{2}$OSe$_{2}$ orders antiferromagnetically (AFM) below $\sim$ 90 K \cite{Free-2010} and two magnetic structures have been discussed: a 2-$k$ model first proposed for Nd$_{2}$O$_{2}$Fe$_{2}$OSe$_{2}$ \cite{Fuwa-2010} and a collinear model \cite{Ni-2011, Free-2010}. Recent work suggests that the 2-$k$ model is the more likely \cite{McCabe-2014, Gunther-2014, Zhao-2013}. In this magnetic structure (Figure \ref{structures}), both next-nearest-neighbour (nnn) interactions (ferromagnetic $J_{2}$ $\sim$ 95$^{\circ}$ Fe -- Se -- Fe interactions, and AFM $J_{2'}$ 180$^{\circ}$ Fe -- O -- Fe interactions) are fully satisfied. Interestingly, a similar 2-$k$ model, composed of perpendicular AFM chains, has been predicted for the FeAs sheets in iron arsenides and oxyarsenides \cite{Lorenzana-2008, Giovanneti-2011}. The recent studies on La$_{2}$O$_{2}$Fe$_{2}$OSe$_{2}$ indicated the presence of antiphase boundaries or stacking faults perpendicular to the stacking direction and gave evidence for 2D-Ising-like magnetism within these sheets \cite{McCabe-2014}. In the light of this work, it is timely to report a similar analysis on $Ln_{2}$O$_{2}$Fe$_{2}$OSe$_{2}$ ($Ln$ = Ce, Nd) analogues to test the robustness of the 2-$k$ model to changes in Fe -- O bond lengths and to a magnetic moment on the $Ln$ site. We show that neutron powder diffraction (NPD) data are consistent with the proposed 2-$k$ magnetic structure and that much larger magnetic stacking domains are formed with these smaller magnetic lanthanides. In contrast with previous work \cite{Ni-2011}, we show that Ce moments order at low temperatures and that unlike many related oxyarsenide systems, this Ce moment ordering does not cause a reorientation of the Fe$^{2+}$ moments.

\section{2. Experimental details}
5 g each of Ce$_{2}$O$_{2}$Fe$_{2}$OSe$_{2}$ and Nd$_{2}$O$_{2}$Fe$_{2}$OSe$_{2}$ were prepared by solid state reaction of CeO$_{2}$ (Alfa-Aesar, 99.8\%) or Nd$_{2}$O$_{3}$ (Electronic Materials, 99\%), Fe (Aldrich, 99.9\%) and Se (Alfa Aesar, 99.999\%) powders. For Nd$_{2}$O$_{2}$Fe$_{2}$OSe$_{2}$, stoichiometric quantities of these reagents were intimately ground by hand using an agate pestle and mortar. The resulting grey powder was pressed to form several pellets of 5 mm diameter. These were placed inside a quartz tube which was evacuated and sealed. This was then heated slowly to 1000$^{\circ}$C, held at this temperature for 12 hours and then cooled in the furnace. For Ce$_{2}$O$_{2}$Fe$_{2}$OSe$_{2}$, a stoichiometric mixture of reagents were ground together and pressed into pellets which were placed in a quartz tube with an oxygen-getter (Al powder in 10$\%$ excess which was physically isolated from the reagents). This tube was evacuated, sealed, slowly heated to 1000$^{\circ}$C and held at this temperature for 12 hours. The reaction tube was then quenched into a bucket of iced water. Both  Ce$_{2}$O$_{2}$Fe$_{2}$OSe$_{2}$ and Nd$_{2}$O$_{2}$Fe$_{2}$OSe$_{2}$ were formed as black, polycrystalline powders. Preliminary structural characterisation was carried out using a Bruker D8 Advance diffractometer fitted with a LynxEye silicon strip detector (step size 0.021$^{\circ}$) operating in reflection mode with Cu K$\alpha$ radiation. NPD data were collected on the time of flight (TOF) diffractometer WISH on target station 2 at the ISIS spallation neutron source. The powders were placed in cylindrical 6 mm diameter vanadium cans (to heights of $\sim$6 cm). For Ce$_{2}$O$_{2}$Fe$_{2}$OSe$_{2}$, 16 minute (10 $\mu$A h) scans were collected every 4 K on cooling from 120 K to 8 K and a final 25 minute (15 $\mu$A h) scan was collected at 1.5 K. For Nd$_{2}$O$_{2}$Fe$_{2}$OSe$_{2}$, several 16 minute (10 $\mu$A h) scans were collected between 1.5 K and 12 K followed by a further five on warming to 300 K. Rietveld refinements \cite{Rietveld-1969} were performed using Topas Academic software \cite{Coelho-2003, Coelho-2012}. Sequential refinements were controlled using local subroutines. The high d-spacing 59$^{\circ}$ bank of data (10.5 -- 85.0 ms, 1 -- 8 \AA) was used to study the magnetic behaviour of the two materials. The nuclear and magnetic refinements were carried out with separate nuclear and magnetic phases (lattices parameters and scale factor of the magnetic phase were constrained to be multiples of the nuclear phase). Magnetic susceptibility was measured using a Magnetic Properties Measurement System (Quantum Design, MPMS). Field-cooled and zero-field-cooled data were collected on warming from 2 K to 292 K at 5 K intervals in an applied magnetic field of 1000 Oe. Field sweep measurements to $\pm$ 5 $\times$ 10$^{4}$ Oe  were made at 292 K and at 12 K. Scanning electron microscopy for compositional analysis of Ce$_{2}$O$_{2}$Fe$_{2}$OSe$_{2}$ samples used a Hitachi SU-70 FEG electron microscope. Back-scattered electron (BSE) images were collected with a YAG BSE detector to investigate compositional homogeneity and energy-dispersive X-ray analysis (EDX) measurements were made over areas of the sample and at  65 -- 100 points on each sample.

\section{3. Results}
The synthesis method described gave a single-phase sample of  Nd$_{2}$O$_{2}$Fe$_{2}$OSe$_{2}$ by X-ray and neutron analysis (see supplementary material). Quenching the sealed reaction tube of Ce$_{2}$O$_{2}$Fe$_{2}$OSe$_{2}$ from the reaction temperature into iced water gave an almost single-phase sample of Ce$_{2}$O$_{2}$Fe$_{2}$OSe$_{2}$ (NPD data revealed the presence of $\sim$ 3$\%$ by weight of a second phase similar to the main phase which was included in refinements). See supplementary information for further analyses. Rietveld refinements using NPD data for both $Ln$ = Ce, Nd phases are in good agreement with the nuclear structure previously reported \cite{Ni-2011}. Refining site occupancies (with $Ln$ site fully occupied) gave compositions close to stoichiometric (Ce$_{2}$O$_{1.99(1)}$Fe$_{2.00(1)}$O$_{0.99(1)}$Se$_{2.0(1)}$ and Nd$_{2}$O$_{1.93(1)}$Fe$_{1.96(1)}$O$_{0.913(9)}$Se$_{2.05(1)}$). Site occupancies were fixed at unity for subsequent refinements. These $Ln$ = Ce, Nd phases can be compared with $Ln$ = La, Pr analogues and the expected decrease in unit cell parameters with decreasing $Ln^{3+}$ ionic radii is observed (\ref{structures}).

\begin{figure}[t] 
\includegraphics[width=1.0\linewidth,angle=0.0]{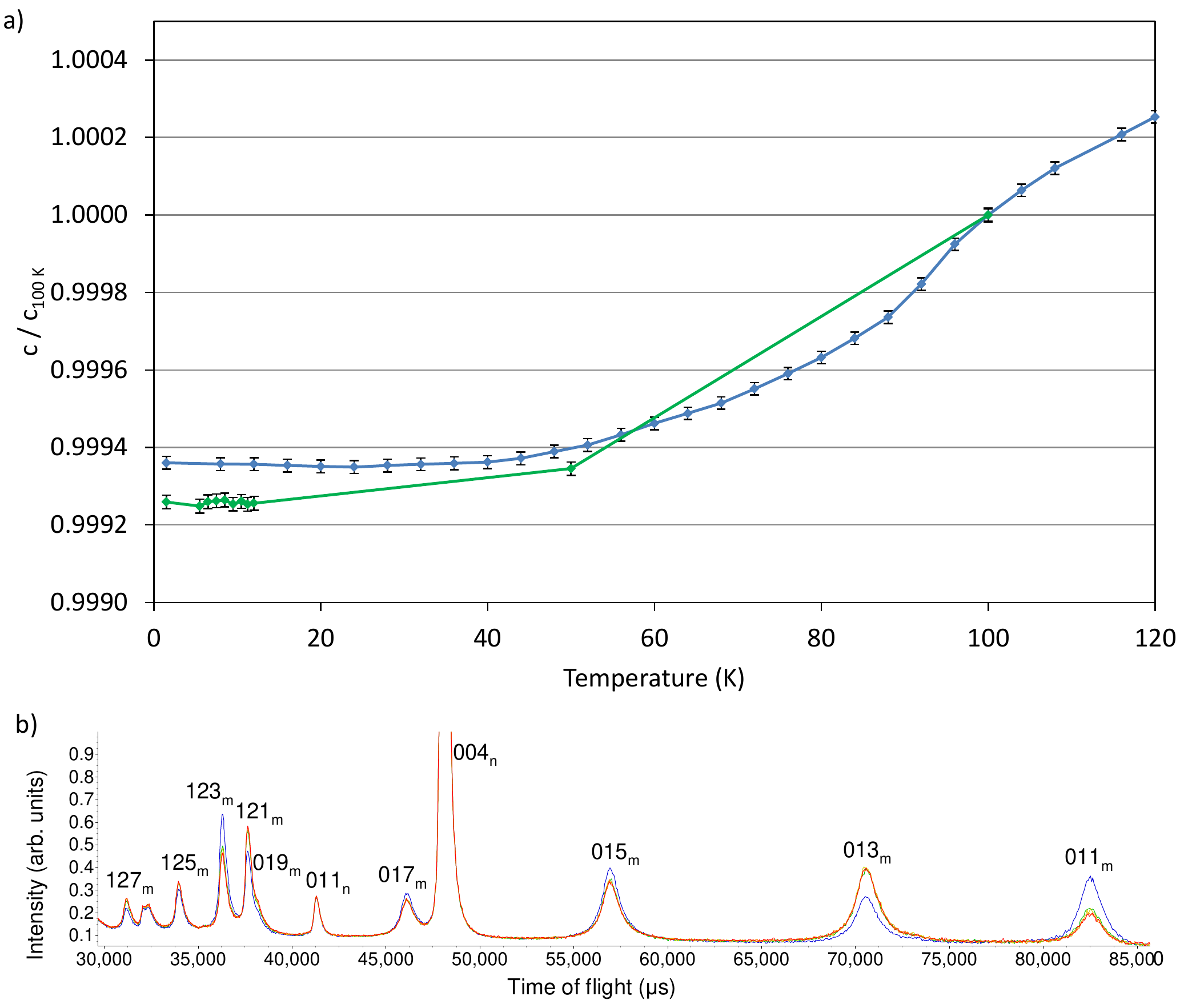}
\caption{[color online]  a) Plots of unit cell parameter $c$ / $c_{100 K}$ for $Ln_{2}$O$_{2}$Fe$_{2}$OSe$_{2}$ ($Ln$ = Ce, Nd) from sequential Rietveld refinements using NPD data, data are shown in blue and green for Ce and Nd analogues, respectively; b) NPD data (59$^{\circ}$ bank from WISH TOF diffractometer) collected for Ce$_{2}$O$_{2}$Fe$_{2}$OSe$_{2}$ at 1.5 K (dark blue), 8 K (green), 12 K (yellow), 16 K (orange) and 20 K (red). $hkl$ indices of key reflections are given and subscripts m and n refer to the magnetic and nuclear phases, respectively.}
\label{vt-diffraction}
\end{figure}

Sequential Rietveld refinements using Ce$_{2}$O$_{2}$Fe$_{2}$OSe$_{2}$ data collected on cooling from 120 K reveal a decrease in unit cell volume on cooling (similar behaviour is observed for Nd$_{2}$O$_{2}$Fe$_{2}$OSe$_{2}$). For both $Ln$ = Ce, Nd analogues, a more rapid decrease of $c$ unit cell parameter is observed below $T_{\mathrm{N}}$ (Figure \ref{vt-diffraction}a) which is similar to that observed for La$_{2}$O$_{2}$Fe$_{2}$OSe$_{2}$ \cite{Free-2010}.

Additional Bragg reflections were observed in NPD data collected for Ce$_{2}$O$_{2}$Fe$_{2}$OSe$_{2}$  below $T_{\mathrm{N}}$ (92.3(2) K) and are consistent with the 2-$k$ magnetic structure reported for La$_{2}$O$_{2}$Fe$_{2}$OSe$_{2}$ \cite{McCabe-2014}. At 96 K, some diffuse scattering is observed at similar d-spacing to the most intense magnetic Bragg reflection (see supplementary material), suggesting some two-dimensional short-range ordering immediately above the transition to a three-dimensional AFM phase. Data were not collected at sufficiently small temperature increments to investigate this for Nd$_{2}$O$_{2}$Fe$_{2}$OSe$_{2}$.

\begin{figure*}[t] 
\includegraphics[width=0.8\linewidth,angle=0.0]{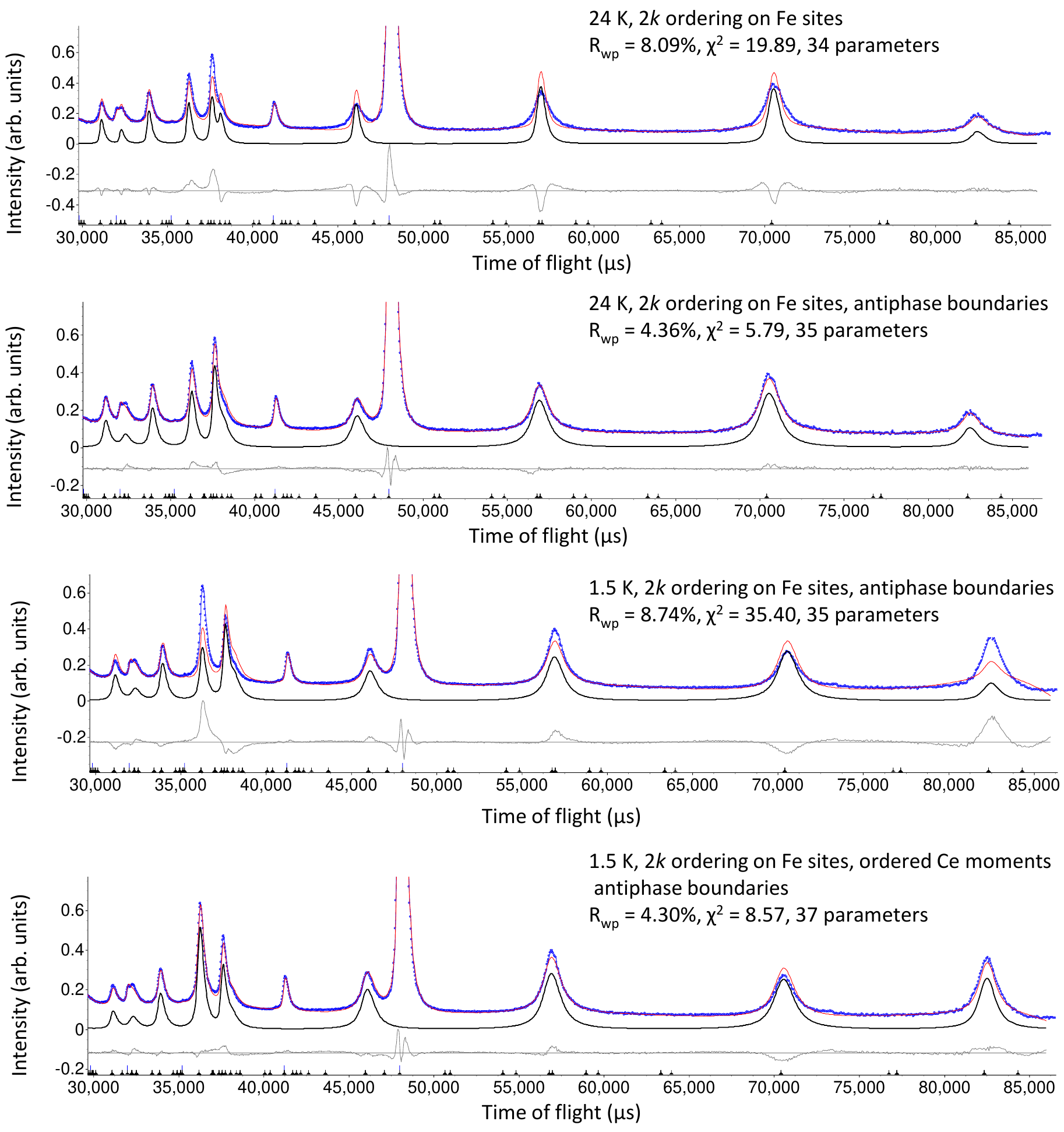}
\caption{[color online]  Rietveld refinement profiles using NPD data (59$^{\circ}$ bank from WISH TOF diffractometer) of Ce$_{2}$O$_{2}$Fe$_{2}$OSe$_{2}$ focusing on the magnetic reflections with 2-$k$ magnetic ordering model on Fe$^{2+}$ sublattice, showing a) refinement at 24 K with the same peak shape for both magnetic and nuclear phases; b) refinement at 24 K including antiphase boundaries for the magnetic phase; c) refinement at 1.5 K with antiphase boundaries for the magnetic phase (only Fe moments ordered) and d) refinement at 1.5 K with antiphase boundaries for the magnetic phase and ordering of both Fe and Ce moments. Observed and calculated (upper) and difference profiles are shown by blue points, red and grey lines, respectively. Magnetic intensity is highlighted by solid black line.}
\label{Ce*boundaries}
\end{figure*}

Significant anisotropic peak broadening of magnetic Bragg reflections was observed for both Ce and Nd phases, similar to that observed for La$_{2}$O$_{2}$Fe$_{2}$OSe$_{2}$ \cite{McCabe-2014}. This could be modelled with an expression for antiphase boundaries perpendicular to the $c$ axis \cite{Her-2007, McCabe-macro} which gave a significantly improved fit to the data (Figure \ref{Ce*boundaries} and supplementary material) and suggested magnetic correlation lengths along $c$, $\xi_{c}$, at 1.5 K of 87(1) {\AA} and 86(2) {\AA} for Ce and Nd analogues, respectively.

For Ce$_{2}$O$_{2}$Fe$_{2}$OSe$_{2}$, the 2-$k$ model with ordered Fe$^{2+}$ moments gives a good fit to NPD data down to $\sim$24 K but a change in relative intensities of magnetic Bragg reflections is observed at lower temperatures (Figure \ref{vt-diffraction}b) and whilst no additional reflections appear, the 24 K model gives a poor fit at 1.5 K (Figure \ref{Ce*boundaries}c).

\begin{figure*}[t] 
\includegraphics[width=1.0\linewidth,angle=0.0]{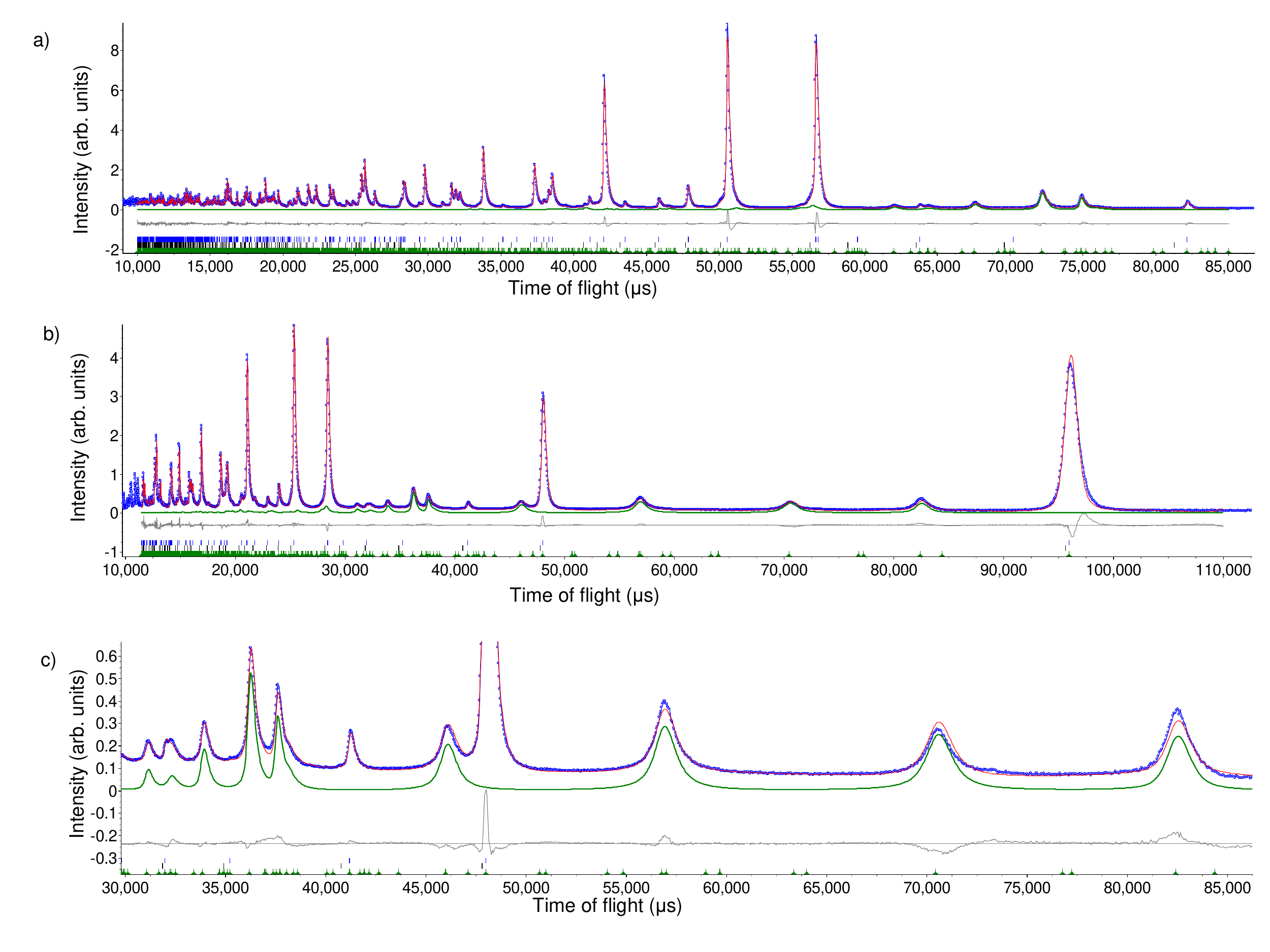}
\caption{[color online]  Rietveld refinement profiles from combined refinement a) using 153$^{\circ}$ bank data and b) 59$^{\circ}$ bank data collected for Ce$_{2}$O$_{2}$Fe$_{2}$OSe$_{2}$ at 1.5 K; c) shows the higher d-spacing region of 59$^{\circ}$ bank data highlighting the magnetic reflections. Observed and calculated (upper) and difference profiles are shown by blue points, red and grey lines, respectively. Magnetic intensity is highlighted by solid green line.}
\label{Ce_1.5K}
\end{figure*}

The low temperature magnetic structure was investigated using the symmetry-adapted mode approach \cite{Campbell-2006} where the magnetic structure is described in terms of the parent crystal structure with a number of symmetry-adapted magnetic ordering modes, or basis vectors, imposed on it. The amplitudes of these modes or basis vectors can be refined to model various magnetic structures. The approach of directly refining the contributions of the different magnetic basis vectors is similar to that originally developed in SARAh \cite{Wills-2000}. Mode inclusion analysis \cite{McCabe-2014, Tuxworth-2013} (see supplementary material for more details) indicated no change in the Fe moment arrangement on cooling. A good fit could, however, be obtained with an ordered moment on Ce sites. The in-plane component of the Ce moments is of the same symmetry as the Fe ordering, but allowing the Ce moments to cant out-of-plane improves the fit further (for refinement using both 153$^{\circ}$ and 59$^{\circ}$ banks of data, R$_{wp}$ decreases from 5.589 $\%$ to 5.127 $\%$ for one additional parameter; see supplementary materials for comparison with Figure \ref{Ce*boundaries}d). A combined refinement using both high resolution data (153$^{\circ}$ bank) and long d-spacing data (59$^{\circ}$ bank) was carried out using data collected at 1.5 K and final refinement profiles are shown in Figure \ref{Ce_1.5K} and refinement details are given in Table \ref{table_Ce_1.5K}.

\begin{table}[ht]
\caption{Details from Rietveld refinement using NPD data collected at 1.5 K for Ce$_{2}$O$_{2}$Fe$_{2}$OSe$_{2}$. The refinement was carried out with the nuclear structure described by space group $I$4/$mmm$ with $a$ = 4.0526(2) \AA and $c$ = 18.440(1) \AA. The magnetic scattering was fitted by a second magnetic-only phase with $a$, $b$ and $c$ unit cell parameters constrained to be twice those of the nuclear phase; R$_{wp}$ = 5.127$\%$, R$_{p}$ = 5.070$\%$ and $\chi^{2}$ = 13.06. }
\centering
\begin{tabular} {c c c c c c c}
\hline\hline
Atom & site & $x$ & $y$ & $z$ & $U_{iso}$ $\times$ 100 (\AA$^{2}$) & moment ($\mu_{B}$)\\
\hline\hline
Ce & 4$e$ & 0.5 & 0.5 & 0.18471(6) & 0.52(4) &  0.89(1) ($xy$)\\
  &   &   &   &   &   &  0.54(1) ($z$)\\
Fe & 4$c$ & 0.5 & 0 & 0 & 0.71(2) & 3.32(1) \\
Se & 4$e$ & 0 & 0 & 0.09755(4) & 0.40(3) &  \\
O1 & 4$d$ & 0.5 & 0 & 0.25 & 0.70(3) &  \\
O2 & 2$b$ & 0.5 & 0.5 & 0 & 0.85(4) &   \\
\hline
\label{table_Ce_1.5K}
\end{tabular}
\end{table}

\begin{figure}[t] 
\includegraphics[width=1.0\linewidth,angle=0.0]{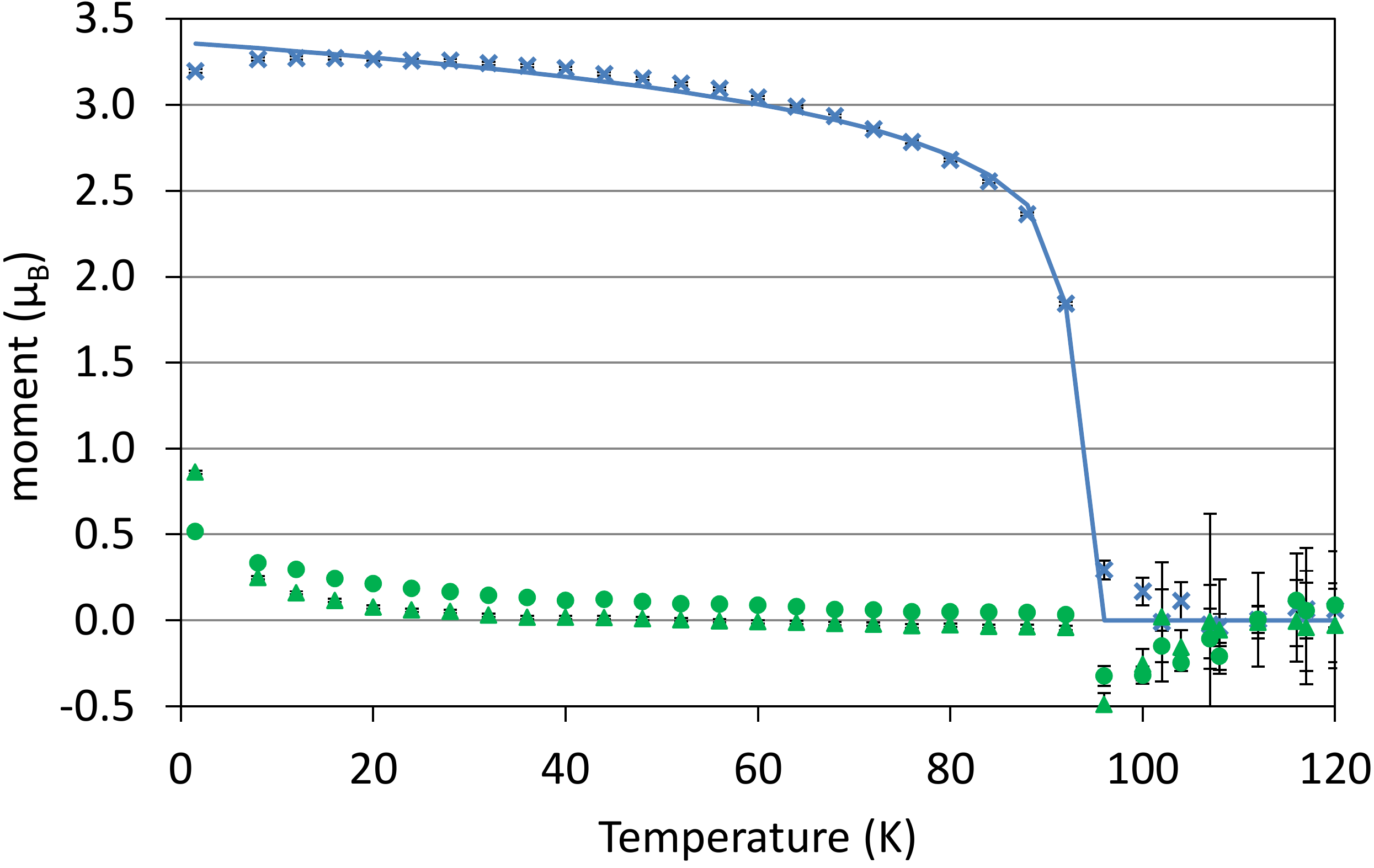}
\caption{[color online]  Evolution of magnetic moment on Fe sites (blue crosses) and on Ce sites in the $ab$ plane (open green triangles) and along $c$ (green crosses) on cooling Ce$_{2}$O$_{2}$Fe$_{2}$OSe$_{2}$; solid blue line shows fit to function $M_{T}$ = $M_{0} (1-({T\over T_{N}}))^{\beta}$ representing critical behaviour of Fe ordering with $M_{0}$ = 3.36(4) $\mu_{B}$, $T_{\mathrm{N}}$ = 92.3(2) K and $\beta_{Fe}$ = 0.11(1).}
\label{critical}
\end{figure}

Sequential Rietveld refinements carried out using only data from the 59$^{\circ}$ bank (which is the more sensitive to magnetic ordering) collected on cooling Ce$_{2}$O$_{2}$Fe$_{2}$OSe$_{2}$ indicate no change in the magnetic correlation length along $c$ and the onset of magnetic order on the Fe sites in the Ce phase can be described by critical behaviour with the exponent $\beta_{Fe}$ = 0.11(1) (Figure \ref{critical}). Moments on Ce sites remain close to zero until $T$ $\leq$ 16 K.

Magnetic susceptibility measurements on both Ce$_{2}$O$_{2}$Fe$_{2}$OSe$_{2}$ and Nd$_{2}$O$_{2}$Fe$_{2}$OSe$_{2}$ are included as supplementary information. They were consistent with those reported by Ni et al \cite{Ni-2011}, showing a slight cusp in susceptibility at $\sim$90 K which is attributed to long range magnetic ordering of the Fe sublattice.

\section{4. Discussion}
The difficulty in preparing a single phase sample of Ce$_{2}$O$_{2}$Fe$_{2}$OSe$_{2}$ may be due to the variable oxidation state of cerium, with both +3 and +4 oxidation states accessible. The secondary phase present in initial syntheses has a smaller unit cell than the almost single-phase material and is thought to contain some Ce$^{4+}$ and vacancies on the Fe$^{2+}$ sites (although we cannot rule out the possibility of some excess oxide ions on interstitial sites). This is similar to reports on the Ce-Cu-O-S system for which both Ce$^{4+}$Cu$_{1-x}$OS and the stoichiometric Ce$^{3+}$CuOS can be prepared \cite{Pitcher-2009}. Based on other $Ln$ systems, the unit cell parameters for the single-phase sample of Ce$_{2}$O$_{2}$Fe$_{2}$OSe$_{2}$ are consistent with those expected for Ce$^{3+}$ (see Figure 1), suggesting that this phase is close to stoichiometric. Site occupancies from Rietveld analysis of the NPD data are consistent with this.

Variable temperature NPD data show that both $Ln$ = Ce, Nd phases remain tetragonal at low temperatures, unlike $Ln_{2}$O$_{2}$Mn$_{2}$OSe$_{2}$ $Ln$ = La, Pr for which low-temperature tetragonal-orthorhombic phase transitions have been reported \cite{Free-2011}. These transitions for the Mn phases are thought to involve the in-plane O(2) site \cite{Free-2011} and may reflect the frustrated Mn -- O(2) -- Mn $J_{2’}$ interactions present in the magnetically ordered phase below $T_{\mathrm{N}}$. This in contrast to the 2-$k$ magnetic structure observed for $Ln_{2}$O$_{2}$Fe$_{2}$OSe$_{2}$ ($Ln$ = La, Ce, Nd) with in-plane Fe -- O(2) -- Fe $J_{2’}$ and Fe -- Se -- Fe $J_{2}$ interactions satisfied \cite{McCabe-2014}. The absence of distortions from tetragonal symmetry is consistent with predictions from theoretical work that such a 2-$k$ structure might suppress any lattice distortions \cite{Lorenzana-2008}. The contraction in unit cell parameter $c$ at the onset of AFM ordering is similar for all $Ln_{2}$O$_{2}$Fe$_{2}$OSe$_{2}$ analogues (Figure \ref{vt-diffraction}a) and less pronounced than for  Mn-containing materials. 

The observation of the same 2-$k$ magnetic structure across the series $Ln_{2}$O$_{2}$Fe$_{2}$OSe$_{2}$ ($Ln$ = La, Ce, Nd) suggests that this ordering pattern is fairly robust with respect to chemical pressure (the $\sim4\%$ decrease in $Ln$ ionic radii from $Ln$ = La to $Ln$ = Ce \cite{Shannon-1976} results in a $\sim4\%$ decrease in unit cell volume \cite{Ni-2011}, with a relatively larger contraction in the $ab$ plane). This is consistent with the low level of frustration in this 2-$k$ magnetic structure, with both nnn $J_{2}$ and $J_{2'}$ interactions satisfied. We note that the magnetic correlation lengths $\xi_{c}$ for Ce$_{2}$O$_{2}$Fe$_{2}$OSe$_{2}$  and Nd$_{2}$O$_{2}$Fe$_{2}$OSe$_{2}$  are almost twice that found for La$_{2}$O$_{2}$Fe$_{2}$OSe$_{2}$  ($\xi_{c}$ = 45(3) \AA  \cite{McCabe-2014}). The greater magnetic correlation length along $c$ in these $Ln$ = Ce, Nd analogues might be due to their decreased interlayer separation.

The ordered moments on the Fe sites reported here for $Ln_{2}$O$_{2}$Fe$_{2}$OSe$_{2}$ (3.32(1) $\mu_{B}$ and 3.18(1) $\mu_{B}$ at 1.5 K for $Ln$ = Ce, Nd, respectively) are larger than those reported previously (2.23(3) $\mu_{B}$ per Fe site at 5 K in Pr$_{2}$O$_{2}$Fe$_{2}$OSe$_{2}$ \cite{Ni-2011}) presumably as a result of improved fitting of the anisotropically broadened of magnetic Bragg reflections and are very similar to that found for La$_{2}$O$_{2}$Fe$_{2}$OSe$_{2}$ (3.50(5) $\mu_{B}$ at 2 K) \cite{McCabe-2014}. The evolution of the Fe moment on cooling observed in Ce$_{2}$O$_{2}$Fe$_{2}$OSe$_{2}$ (and the beta exponent $\beta_{Fe}$ obtained) is very similar to that reported for La$_{2}$O$_{2}$Fe$_{2}$OSe$_{2}$ ($\beta_{Fe}$ = 0.122(1)) and consistent with 2D-Ising-like behaviour reported for several related systems \cite{McCabe-2014, Fuwa-2010}. The diffuse scattering observed only 4 K above $T_{\mathrm{N}}$ for Ce$_{2}$O$_{2}$Fe$_{2}$OSe$_{2}$ is also consistent with the 2D-like spin fluctuations before the onset of 2D magnetic order. As noted by Ni et al, the lack of a clear maximum in susceptibility data for $Ln$ = Ce, Nd analogues (and the small entropy release observed in their careful heat capacity measurements) again suggests a 2D to 3D magnetic transition transition \cite{Ni-2011}.

Ordering of Ce moments at low temperature in Ce$_{2}$O$_{2}$Fe$_{2}$OSe$_{2}$ has not been reported previously. This may reflect the difficulty in preparing stoichiometric samples of Ce-containing transition metal oxychalcogenides which can easily contain transition metal vacancies and diamagnetic Ce$^{4+}$ cations, or the difficulty in detecting small ordered moments (1.08(1) $\mu_{B}$ per Ce site observed here) which only develop at very low temperatures.  The Ce moment obtained is consistent with that  expected for a Ce$^{3+}$ doublet ground state (1 $\mu_{B}$) \cite{Chi-2008} and similar to that observed in related Ce$_{2}$O$_{2}$FeSe$_{2}$ (1.14(4) $\mu_{B}$ at 4 K) \cite{McCabe-2014-Ce2O2FeSe2} and CeFeAsO (0.83(2) $\mu_{B}$ at 1.7 K) \cite{Zhao-2008}. 

Below $T_{\mathrm{N,Ce}}$ Ce moments have a similar inplane arrangement to the Fe moments (Figure \ref{structures}), although a large component is directed along $c$.  This is in contrast with related CeMnAsO in which Ce moments lie within the $ab$ plane \cite{Lee-2012, Tsukamoto-2011}, consistent with the easy-axis along $x$ proposed for Ce$^{3+}$ sites in fluorite-like [Ce$_{2}$O$_{2}$]$^{2+}$ layers in related orthorhombic materials \cite{Gornostaeva-2013}. In CeFeAsO, Ce moments are predominantly within the $ab$ plane but a small out-of-plane component has been proposed \cite{Zhao-2008}. This canting of Ce moments away from the easy-axis may indicate some Fe -- Ce coupling in these iron-based mixed anion systems. We find no evidence of Fe moment reorientation at the onset of Ce$^{3+}$ moment ordering in Ce$_{2}$O$_{2}$Fe$_{2}$OSe$_{2}$ (which we expect our analysis to be sensitive to, given the relatively large moments on Fe sites). This is in contrast to CeMnAsO \cite{Lee-2012} and CeFeAsO \cite{Zhang-2013} where some reorientation of transition metal spins has been reported at $T_{\mathrm{N,Ce}}$. The lack of any spin reorientation in Ce$_{2}$O$_{2}$Fe$_{2}$OSe$_{2}$ may reflect the Ising-like nature of the Fe ordering, with the local anisotropy of Fe directing moments oriented along Fe -- O(2) bonds and parallel to the expected Ce easy axis. Analysis of the magnetic order in La$_{2}$O$_{2}$Fe$_{2}$OSe$_{2}$ by G$\ddot{u}$nther et al \cite{Gunther-2014} has shown that the 2-$k$ magnetic ordering on the Fe sublattice gives rise to two inequivalent $Ln$ sites. Refinements allowing inequivalent Ce sites did not improve the fit and our data may not be sensitive to this.

\section{5. Conclusions}
In conclusion, we have shown that NPD data for $Ln_{2}$O$_{2}$Fe$_{2}$OSe$_{2}$ ($Ln$ = Ce, Nd) are consistent with the 2-$k$ magnetic structure and, like the La analogue, ordering of the Fe$^{2+}$ moments is characterised by 2D-Ising-like spin fluctuations around the critical point. Below 16 K, Ce moments order with a similar 2-$k$ arrangement but with a large out-of-plane component. The lack of any reorientation of Fe moments at the onset of Ce spin order may reflect the Ising-like nature of the Fe moments. Canting of Ce moments away from their easy-axis may indicate some Fe -- Ce coupling.

See Supplemental Material at [URL will be inserted by publisher] for further details regarding synthesis, analysis of diffraction data (including mode inclusion analysis) as well as magnetic susceptibility and field sweep measurements..

We acknowledge STFC, EPSRC (EP/J011533/1) for funding. We thank Dmitry Khalyavin (ISIS), Budhika Mendis (Durham) and Leon Bowen (Durham) for assistance and Chris Stock, Efrain Rodriguez and Mark Green for helpful discussions.

\end{document}